\documentstyle[aps,prl,multicol]{revtex}  
\begin{document}
\title{\vskip -0.4truecm
\hfill \draft{CINVESTAV-FIS/97-13} \\
\hfill \draft{ICN-UNAM-97-13} \\
\vskip 0.5truecm
Universal Scaling Properties of Superconductors in Magnetic Fields
}
\author{Denjoe O'Connor}
\address{Departamento de F\'\i sica,\\
Cinvestav, Apartado Postal 14-740, \\
M\'exico D. F. 07000}
\author{C.R. Stephens}
\address{Instituto de Ciencias Nucleares,\\
UNAM, Circuito Exterior, Apartado Postal 70-543,\\
M\'exico D.F. 04510.}
\date{\today}
\maketitle

\begin{abstract}
Based on renormalization group arguments we establish that for a
superconductor 
in the presence of a weak external magnetic field, $B$, the dependence 
on $B$ and the deviation from the critical temperature, $\tau$, 
of a thermodynamic quantity, $P$, takes the scaling form 
$P=t^{\theta}X({B\over\Phi_0}\tau^{-2\nu},q\tau^{-\nu\omega_e})$, 
where $\theta$ and $\nu$ are XY exponents, $q$ is the scaled electromagnetic 
coupling and $\nu\omega_e$ is the associated crossover exponent.
For $q/\tau^{\nu\omega_e}\ll1$, the experimentally accessible region in
high-$T_c$ superconductors, there is a reduction to one-variable scaling
plus small corrections. In this region we find the shift in the 
specific heat maximum is given by $\Delta=x_0{(B/\Phi_0)}^{1/2\nu}$ 
and that the singular part of the free energy at the critical temperature 
takes the form $F_{sing}=c(d){(B/\Phi_0)}^{d/2}$ where $c(d)$ is a universal 
amplitude. A one loop approximation in three dimensions gives
$c(3)\sim0.22$.
The results presented here should have equal applicability to 
the nematic to smectic-A transition.
\end{abstract} 

\begin{multicols}{2}

Since the discovery of high-$T_c$ superconductivity there has 
been a significant
increase in interest in the critical properties of superconductors. 
As one of the principal experimental probes 
for the analysis of high-$T_c$ materials involves 
their study in external magnetic fields it is important 
to have reliable and robust theoretical predictions for such a setting. 

The formal analogy between a superconductor in an external magnetic field and 
a system constrained by finite size is well known \cite{analogy}. 
In \cite{ourpap} we established that correlation functions 
of such a superconductor exhibit a 
scaling form identical to that of a near critical system in a confined 
geometry of size $L$, with the combination $(B/\Phi_0)^{-1/2}$, where $B$ 
is the magnetic field and $\Phi_0={\pi \hbar c/ e}$ is the unit 
flux quantum, playing the role of $L$. A crucial
ingredient in our analysis was the assumption of a non-trivial, 
infrared stable 
renormalization group (RG) fixed point. In conventional superconductors
such a fixed point is experimentally inaccessible. Additionally, there are
theoretical arguments \cite{hlma} that cast doubt on the existence of 
such a fixed point except in the artificial case where the number of 
components of the order parameter is large. For type-II superconductors
there is some evidence \cite{dashal} that very close to $T_c$
the relevance of magnetic field fluctuations causes a crossover
to a stable ``charged-XY'' fixed point with inverted temperature axis.
Once again in the case of conventional superconductors this crossover would 
be unobservable.
 
On a more pragmatic note, experimentally there is now ample evidence 
\cite{e1,e2,e3} that for high-$T_c$ superconductors, 
for $\vert T-T_c\vert\leq 10K$, 
the zero field transition exhibits 
critical behaviour characteristic of a three dimensional $XY$-model.
This fact, together with the finite size analogy, 
has been utilized \cite{Inderheesetal} to collapse experimental 
specific heat data using a ``finite size'' scaling ansatz of the form
$P=\tau^{\theta}X({B\over\Phi_0}\tau^{-2\nu})$, 
where $\tau$ is a linear deviation from the critical temperature 
and $\nu$ and $\theta$ are $XY$ exponents, with $\nu\sim0.67$ 
for a three dimensional sample. In a recent letter Lawrie \cite{Lawrie} 
presented two-loop arguments establishing that to this order, 
when fluctuations in the electromagnetic field are neglected, 
the two-point function 
of the order parameter in a magnetic field exhibits such a scaling form.

The neutral-$XY$ fixed point where $e^*=0$ is unstable 
with respect to a small perturbation of the electromagnetic coupling.
Hence, one expects a region wherein electromagnetic
fluctuations can be treated as a relevant perturbation, 
in a similar fashion to temperature, thus leading to a two-variable 
scaling formulation.   
For high-$T_c$ superconductors, where the Ginzburg-Landau parameter,
$\kappa$, is large (e.g. for $YBa_2Cu_3O_{7-y}$, $\kappa\sim 87$ 
\cite{Blatterreview})
the corrections due to electromagnetic fluctuations,
although relevant, are very small except in a region
asymptotically close to $T_c$ and
can be taken into account as corrections to scaling around
the single variable scaling form associated with the 
neutral-$XY$ fixed point. 
Experimental evidence certainly seems
to accord with this picture.

In this letter we establish using RG techniques that the appropriate scaling 
form for a general thermodynamic quantity, $P$, in a magnetic field takes the
form 
$P=t^{\theta}X({B\over\Phi_0}\tau^{-2\nu},q\tau^{-\nu\omega_e})$.
In the case where electromagnetic fluctuations are weak, 
such as high-$T_c$ superconductors, we show that this reduces 
to a one-variable scaling form plus small corrections to scaling
relative to the $XY$ fixed point. In addition, our analysis provides
several new theoretical predictions which should be testable as experiments 
on high-$T_c$ 
superconductors become more refined in the vicinity of the critical
point. The derivation elucidates the assumptions on which such a 
scaling form relies, in particular the result is quite robust.

Our starting point is the Landau-Ginzburg-Wilson 
Hamiltonian for a superconductor, based on the Cooper pair 
density as order parameter, given by
\begin {eqnarray} 
H[\varphi ]&= \int d^d{\bf 
x}\bigg(\vert(\nabla-{2ie\over\hbar c} {\bf A})\varphi\vert^2 
+\ r_0 |\varphi|^2 
+\frac{\lambda_0}{6}|\varphi|^4 
\nonumber \\
&+\frac{1}{8\pi}(\nabla\times{\bf A})^2
-{1\over 4\pi}{\bf B\cdot\nabla\times A}\bigg),
\label{LGW} 
\end{eqnarray} 
Note that with the substitutions, $r_0={2\alpha m\over\hbar^2}$, 
$\lambda_0={12\beta m^2\over\hbar^4}$ and $\varphi={\hbar\over\sqrt{2m}}\psi$
we recast  Eq. (\ref{LGW}) into conventional form 
(as found in e.g. Blatter et al.\cite{Blatterreview}). The only temperature 
dependent parameter is $\alpha$ (and hence $r_0$), which is 
assumed to depend linearly on $T$. 
Though high-$T_c$ superconductors are highly anisotropic, if the anisotropy 
is not so strong as to invalidate a continuous LGW formulation, the 
asymmetric results can be obtained from 
the symmetric ones by the rescaling methods of Blatter, Gershkenbein and Larkin
\cite{BlatterGesL}, hence we concentrate on the symmetric case.

In the case of a superconductor in a unidirectional magnetic field, $B$, 
fluctuations of 
$\varphi$ induce a renormalization of $B$ even if 
electromagnetic fluctuations are ignored. This renormalization becomes 
ultraviolet divergent at the upper critical dimension and is encoded in the
wavefunction renormalization of the vector potential $\bf A$, i.e. 
${\bf A}=Z^{1/2}_A{\bf A_0}$, (where we use a subscript 
$0$ to indicate a bare or unrenormalized quantity). The electromagnetic 
coupling $e=Z_ee_0$ is similarly renormalized. Gauge invariance as
implemented by the Ward identities relates the two renormalizations, 
$Z_eZ^{1/2}_A=1$ \cite{itzzub}. An immediate consequence is that
$eB$ is an RG invariant. 
Furthermore, from the Hamiltonian (\ref{LGW}) we see that only 
the combination ${B/\Phi_0}$ arises in any correlation function 
or thermodynamic quantity associated with the $\varphi$ field. 
Thus $(\Phi_0/B)^{1/2}$ is a nonlinear scaling field with the 
dimensions of {\it length} which behaves for RG purposes in a fashion very 
similar to the linear size $L$ of a finite or layered system.

We begin by establishing the scaling form for a general 
correlation function. From the Hamiltonian 
(\ref{LGW}) we see that correlation functions 
depend on $r_0$, $\lambda_0$, $e_0$ and the RG invariant $B/\Phi_0$ 
together with a cutoff $\Lambda$.
Defining the renormalized correlation functions $G^{(N,M)}$ by
\begin{eqnarray}
<\varphi^*(1)..\varphi^*(k)\varphi(k+1)..\varphi(N)
\vert\varphi(1)\vert^2..\vert\varphi(M)\vert^2>
\nonumber\\
=G^{(N,M)}(t,\lambda,e,{B\over\Phi_0},\mu,x_1,.,x_N,y_1,.,y_M)
\label{rencorr}
\end{eqnarray}
where $t$ is the renormalized deviation from the zero field critical 
temperature, i.e. $r_0=r_c+Z_{\varphi^2}t$,
the relation between the bare and renormalized correlation functions is
\begin{eqnarray}
G^{\scriptscriptstyle(N,M)}(t,\lambda,{B\over\Phi_0},e,\mu)
=Z_{\varphi}^{\scriptscriptstyle-{N\over2}}Z_{\varphi^2}^{\scriptscriptstyle M}
G^{\scriptscriptstyle(N,M)}_0(r_0,\lambda_0,{B\over\Phi_0},e_0,\Lambda)
\nonumber\end{eqnarray}
where for convenience we have suppressed the dependence on position.
$Z^{1/2}_{\varphi}$ and  $Z_{\varphi^2}$ are the field and 
composite operator renormalization factors,
$Z_{\lambda}\lambda=\lambda_0$ and $\mu$ is an arbitrary renormalization scale.
Here we take $\lambda$ and $e$ to be the dimensionless renormalized couplings. 

The Hamiltonian (\ref{LGW}) defines a renormalizable 
field theory for any dimension $d\leq4$. As ultraviolet divergences are 
unaffected by the presence of an external $B$ field the
$Z_i$ obtained via a $B$ independent renormalization scheme, such as minimal 
subtraction, are sufficient to renormalize the theory for
$B\neq0$. The RG equation obeyed by $G^{(N,M)}$ is equivalent 
to the statement that the bare functions are independent of the arbitrary 
renormalization scale chosen to define the renormalized theory; 
hence the renormalized
correlation functions obey the equation
\begin{eqnarray}
\bigg(\mu{\partial\over\partial\mu}
+\gamma_{\varphi^2}t{\partial\over\partial t}
+\beta_e{\partial\over\partial e}
+\beta_{\lambda}{\partial\over\partial \lambda}\bigg)G^{(N,M)}
\nonumber\\
-(M\gamma_{\varphi^2}+\frac{N}{2}\gamma_{\varphi})G^{(N,M)}=0,
\label{RGofG}
\end{eqnarray}
where $\beta_{e}=\mu {de/d\mu}$, 
$\beta_{\lambda}=\mu{d\lambda/ d\mu}$ and 
$\gamma_{\varphi}=\mu {d\ln Z_{\varphi}/ d\mu}$ and 
$\gamma_{\varphi^2}=-\mu {d\ln Z_{\varphi^2}/d\mu}$ are the
anomalous dimensions of $\varphi$ and $\varphi^2$ respectively.
The solution of Eq. (\ref{RGofG}) is
\begin{eqnarray}
G^{(N,M)}(t,\lambda,e,{B\over\Phi_0},\mu)={\rm e}^{\int_1^\rho\big(\frac{N}{2}
\gamma_{\varphi}(x)+M\gamma_{\varphi^2}(x)\big){dx\over x}}\nonumber\\
\times G^{(N,M)}(t(\rho),\lambda(\rho),e(\rho),{B\over\Phi_0},\rho\mu).
\label{RGofGsoln}
\end{eqnarray}
The absence of a term of the form $\beta_B{\partial/\partial B}$ in
(\ref{RGofG}) is due to the fact that $B$ only appears in the RG 
invariant combination $B/\Phi_0$ and is directly analogous to the
absence of a term $\beta_L{\partial /\partial L}$ in the RG equation 
for a finite system. The latter is a crucial element in the 
proof of finite size scaling \cite{Brezin}.

To proceed further we need to make some assumption about the fixed points
of the model. The Landau-Ginzburg model under discussion has several 
potential candidates. In \cite{ourpap} we assumed an infrared
stable fixed point, $e=e^*$, $\lambda=\lambda^*$. 
By going to a ``dual'' formulation there is some theoretical evidence 
\cite{dashal} that
for type-II superconductors the zero field transition is associated with a 
fixed point of this type which is in the universality class of the 
charged-$XY$-model with inverted temperature axis. 
This fixed point is associated with only one relevant 
direction, $t$, and in its vicinity scaling functions depend 
on only one scaling
variable. In the case of superconductivity there is at present no experimental
evidence for such a fixed point. Another candidate is the Gaussian
fixed point $\lambda^*=0$, $e^*=0$, however, a scaling formulation  based on an
expansion around this fixed point is in disagreement with experimental 
results for high-$T_c$ superconductors. 

Given that there is compelling experimental evidence that the 
$XY$ fixed point is relevant to critical fluctuations in 
high-$T_c$ superconductors we will now concentrate our attention on 
a scaling description based on the fixed point:
$e^*=0$, $\lambda=\lambda^*$ where $\lambda^*$ is 
the Wilson-Fisher fixed point coupling for
an $O(2)$ model. 
One might worry that at such a fixed point, since $e^*=0$,  
the dependence on $B$ would drop out. This is not the
case since $eB$ is an RG invariant and if it is non-zero at
the beginning of an RG flow it is non-zero at the end.

There is an essential difference between this fixed point in the context of a 
superconductor and the standard XY case since in the scaling 
neighbourhood of this 
neutral-$XY$ fixed point, as can be seen from the work of 
Halperin, Lubensky and Ma \cite{hlma}, there are two relevant variables, 
$t$ and $e$, and therefore one has generically two-variable scaling 
functions, with $t$ being the leading relevant perturbation.
As $\beta_e(\lambda,e)=-\omega_e(\lambda,e)e$ linearization about
$e^*=0$ ensures that there is no mixing of $e$ and $\lambda$ 
in the relevant direction, though the irrelevant direction is 
a linear combination of $e$ and $\lambda$ \cite{hlma}. 
For $d=3$ estimates give $\nu\sim 0.67$ and $\omega_e\sim 1$.

Since the RG equations for the two relevant variables are
\begin{eqnarray}
\rho{de\over d\rho}=-\omega_e(\lambda,e)e\qquad
\hbox{and }
\rho{dt\over d\rho}=\gamma_{\varphi^2}(e,\lambda)t,
\label{relvars}
\end{eqnarray}
in the scaling neighbourhood of the fixed point we have 
$\gamma_{\varphi^2}(\lambda^*,0)=2-1/\nu$, where $\nu$ is the 
XY correlation length exponent 
and $\omega_e(\lambda^*,0)=\omega_e$ is the scaling dimension of the charge.
Integrating (\ref{relvars}) yields 
$t(\rho)=t\rho^{2-1/\nu}$ and $e(\rho)=e\rho^{-\omega_e}$.
The general scaling form of (\ref{RGofGsoln}) then becomes
\begin{eqnarray}
G^{\scriptscriptstyle(N,M)}={(\rho\mu)}^{\theta_{\scriptscriptstyle{N,M}}}
{\cal G}^{\scriptscriptstyle(N,M)}({B\over\Phi_0{(\rho\mu)}^2},
{\tau\over{(\rho\mu)}^{1/\nu}},{q\over{(\rho\mu)}^{\omega_e}})
\label{scalingGnm}
\end{eqnarray}
where $\theta_{N,M}={N\over2}(\nu d-\gamma)+M(\nu d-1)$,
$\tau=t\mu^{1/\nu-2}$, $q=e\mu^{\omega_e}$ and 
$G^{(N,M)}$ has been rescaled by a factor 
$\mu^{-{N\over2}\eta-M(2-{1\over\nu})}$. 
Since the co-ordinates $x_i$ do not get renormalized they enter
the dimensionless functions ${\cal G}^{(N,M)}$ in the combination 
$x_i\rho\mu$. Eq. (\ref{scalingGnm}) implies $G^{(N,M)}$ 
is a homogeneous function, therefore the 
variable $\rho$ can be used to eliminate one of the arguments.
We choose to eliminate $\rho$ by setting 
${\tau/{(\rho\mu)}^{1/\nu}}=1$. 
The general scaling form in the vicinity of the 
neutral-$XY$ fixed point is then
\begin{eqnarray}
G^{(N,M)}=\tau^{\theta_{N,M}}
{\cal G}^{(N,M)}(x,y)
\label{scform}
\end{eqnarray}
where the scaling variables are $x={B/\Phi_0{\tau}^{2\nu}}$ and
$y={q/{\tau}^{\nu\omega_e}}$ and ${\cal G}^{(N,M)}$ is a universal
scaling function.
A single-variable scaling function is recovered in the 
limit $q\rightarrow0$, or
alternatively in the limit $T\rightarrow T_c$ if there exists 
a non-trivial stable infrared fixed point.

The free energy takes the two-variable
scaling form 
\begin{eqnarray}
F_{sing}=\tau^{2-\alpha}{\cal F}(x,y)\label{freeen}
\end{eqnarray}
with $x$ and $y$ as above.
When the temperature is precisely tuned to the critical temperature
it is no longer appropriate to eliminate $\rho$ in terms of the 
temperature variable $\tau$ but instead one can use the condition 
${B/\Phi_0{(\rho\mu)}^2}=1$. 
The dependence on the external field of the free energy 
then takes the simplified form
\begin{eqnarray}
F_{sing}={\left({B\over\Phi_0}\right)}^{d/2}
{\cal F}\left(q{\left(\Phi_0\over B\right)}^{\omega_e/2}
\right)
\label{free}
\end{eqnarray}
with a corresponding expression for the specific heat.

We have shown here that generically one expects to see two-variable
scaling functions when electromagnetic fluctuations are taken into account, 
however, one-variable scaling seems to give a very good fit to current 
experimental data. The consistency of these two seemingly contradictory
facts can be seen by making a Taylor expansion of the scaling functions
(\ref{scform}-\ref{free}). 
One finds, for instance, for the correlation functions 
to first order in $e$
\begin{eqnarray}
G^{(N,M)}=\tau^{\theta_{N,M}}
(1+q\tau^{-\omega_e\nu}D(x)){\cal G}^{(N,M)}(x).		
\label{scformred}
\end{eqnarray}
For strongly type-II superconductors the correction to scaling factor, even
though it is associated with a relevant operator, will be very small unless 
one gets very close to the critical point, and hence a one-variable scaling
form associated with the $XY$ fixed point should offer a very good 
description. Naturally it would be very interesting
to see if such corrections to scaling due to electromagnetic fluctuations
could be observed experimentally. 

With the scaling formulation given above and the analogy with finite size 
scaling there are several interesting
corollaries regarding critical temperature shifts and rounding. 
For $d<4$ one does not expect any critical divergence for $B\neq0$,
hence any shifts will be associated with a pseudo-critical temperature
$T_p(B)$, which we take to be the point at which the specific heat is a 
maximum, i.e. $\partial C/\partial\tau\vert_{\scriptscriptstyle T=T_p(B)}=0$. 
Given that the specific heat is proportional to 
$\int d^dx G^{(0,2)}$, then $\Gamma^{(0,3)}=\partial C/\partial\tau$
has the scaling form
\begin{eqnarray}
\Gamma^{(0,3)}(\tau,\lambda,e,{B\over\Phi_0})=\tau^{d\nu-3}{\cal M}(x,y).
\label{gthree}
\end{eqnarray}
The shift in temperature of the specific heat maximum
$\Delta\propto T_p(B)-T_c(0)$ is given by the zero of the 
universal scaling function $\cal M$, i.e.
${\cal M} (x,y)=0$.
Once $y=q/\Delta^{\nu\omega_e}\ll 1$ we can set $q=0$ in (\ref{gthree}),
a good approximation experimentally, 
(corrections to scaling can of course be added).
Denoting this universal zero, $x_0$, one finds
\begin{eqnarray}
\Delta=x_0\left({B\over\Phi_0}\right)^{1/2\nu}
\end{eqnarray}
which is identical to the corresponding formula familiar in finite size scaling
\cite{BarberandFisher}.
A similar expression for the specific heat rounding follows from the 
standard argument that finite size effects will 
start to become important when $
\xi_0\sim ({B/\Phi_0})^{-1/2}$, where $\xi_0$ is the zero field
coherence length. 

The above general argument can also be used to derive 
in the present context the analog of phenomenological renormalization 
\cite{night}. Defining the coherence length via the second moment of 
the two-point correlation function leads to
\begin{equation}
{\xi(B)\over\xi_0}={\cal R}\left({B\xi^2_0\over\Phi_0}\right)
\end{equation}
where $\cal R$ is a universal scaling function and once again we are
neglecting electromagnetic corrections to scaling.
Defining a scale transformation $B\rightarrow B'$, 
$t\rightarrow t'$ via
$B\xi^2(t,0)=B'\xi^2(t',0)$ leads to
\begin{equation}
\xi^2\left(t,{B\over\Phi_0}\right)/\xi^2\left(t',{B'\over\Phi_0}\right)=
B'/B
\end{equation}
which represents an exact RG transformation. $T=T_c(0)$ is a fixed point of
this equation and the critical exponent $\nu$ can be obtained by linearizing 
around this fixed point to find
\begin{equation}
\left(1+{1\over\nu}\right)\ln{B\over B'}=\ln{{\dot\xi}(0,{B'\over\Phi_0})
\over{\dot\xi}(0,{B\over\Phi_0})}.
\end{equation}

In the approximation where electromagnetic fluctuations can be neglected, 
or if a charged-$XY$ fixed point exists, the above scaling formulation also 
leads to many new universal critical-point ratios. 
For instance, the ratio $\xi^2B/\Phi_0$, 
evaluated at the critical temperature, 
should be a universal number. Also, the singular part of the free energy 
at $T=T_c(0)$ will have the form 
$F_{sing}= c(d)(B/\Phi_0)^{d/2}$ where
$c(d)$ is a universal number analogous to the Casimir amplitude familiar in 
finite size scaling \cite{krechbook}. A one loop approximation gives
\begin{equation}
c(d)=-\Gamma(1-{d\over2})\zeta(1-{d\over2},{1\over2})
\label{cpfd}
\end{equation} 
which for $d=3$ gives $c(3)\sim0.22$.

To conclude: we have derived using RG methods the scaling
behaviour of thermodynamic functions in the critical region for
a superconductor in a magnetic field. We find that in the 
neighborhood of the $XY$ fixed point the relevance of electromagnetic
fluctuations leads to a two-variable scaling formulation. 
For $q/\tau^{\nu\omega_e}\ll 1$, the experimentally accessible region
for high-$T_c$ superconductors, this two-variable formulation reduces to 
a one-variable form plus corrections to scaling due to the electromagnetic
fluctuations.

We believe that various 
predictions we have made here should be experimentally verifiable in 
high-$T_c$ superconductors. In particular, the analogy 
with finite size scaling is robust and it should be possible
to determine in exactly what regions the data collapses onto a 
one-variable scaling formulation and to what extent corrections to scaling
can describe any deviations. It should also be possible to measure
some of the universal critical point ratios that arise naturally in the 
present scaling formulation and the predicted 
shift of the specific heat maximum. It would be interesting to extend
the present approach to include
dynamics, anisotropy and/or disorder. The conclusions of this letter
should also be of relevance to the nematic to smectic-A phase transition 
in liquid crystals which is believed to be in the same universality class.

This work was supported in part by Conacyt under grant number 211085-5-0118PE.

\end{multicols}  
\end{document}